\documentstyle [12pt,a4p,epsfig,amsmath,multicol]{article}
\begin{document}
\vspace*{0.6cm}

\begin{center} 
{\normalsize\bf Remarks on: `Theory of Neutrino Oscillations' (hep-ph/0311241) by
   C.Giunti, the comments by L.B.Okun et al in hep-ph/0312151, and
   Giunti's reply in hep-ph/0312180}
\end{center}
\vspace*{0.6cm}
\centerline{\footnotesize J.H.Field}
\baselineskip=13pt
\centerline{\footnotesize\it D\'{e}partement de Physique Nucl\'{e}aire et 
 Corpusculaire, Universit\'{e} de Gen\`{e}ve}
\baselineskip=12pt
\centerline{\footnotesize\it 24, quai Ernest-Ansermet CH-1211Gen\`{e}ve 4. }
\centerline{\footnotesize E-mail: john.field@cern.ch}
\baselineskip=13pt
 
\vspace*{0.9cm}
\abstract{As pointed out previously, some of the
 hypotheses used in the standard discussion
  of the quantum mechanics neutrino oscillations, recently summarised in the
  paper `Theory of Neutrino Oscillations' by C.Giunti, lack
  any physical foundation and lead to neglect of both a factor
  of two correction in the contribution of neutrino propagation
  to the oscillation phase and an important contribution to the
  latter from the propagator of the source particle.}
\vspace*{0.9cm}
\normalsize\baselineskip=15pt
\setcounter{footnote}{0}
\renewcommand{\thefootnote}{\alph{footnote}}
\newline
{\it Keywords ;} Quantum Mechanics,
Neutrino Oscillations.
\newline

\vspace*{0.4cm}

  
   C.Giunti has performed a valuable service in making a clear and concise 
   statement in Ref.~\cite{Giunti1} (TNO) of the basic physical 
  assumptions underlying the standard quantum mechanical treatment of
  neutrino oscillations. Of particular importance are the assumptions
  A2 and A3 of TNO which are the essential ones in order to derive the
  standard result for the oscillation phase. However, A2, as written, is either
  factually incorrect, or, on the most generous interpretation highly likely
  to be misinterpreted. Indeed, it is the incorrect interpretation that is
  used in the standard derivation. In fact the initial and final states
  in the amplitudes of any standard model process are, uniquely, the mass
  eigenstates $| \nu_k \rangle$ not the `neutrino flavour eigenstate'
  $| \nu_{\alpha} \rangle$ defined in Eqn(1) of TNO.
  Since the MNS matrix element, $U_{\alpha k}$,
  necessarily appears in the amplitude when the state  $| \nu_k \rangle$
   is created or destroyed, in association with a charged lepton 
   with label  $\alpha =  e, \mu , \tau$, one can introduce, formally,
     the state
    $| \nu_{\alpha} \rangle$ in order to write, in a 
    compact way, a general expression for the charged weak current or the
    corresponding Lagrangian, always bearing in mind, however, that it has no actual
  physical significance, as the state does not appear in the amplitude of any 
   standard model process. Now what is done in the standard derivation
   is to assume that this formal, physically non-existent state 
   \underline{is actually created by the neutrino production
    process at some fixed time}. Since the detection process
    evidently occurs at another fixed time, the implicit assumption
   is made that all the mass eigenstates have the same velocity, which
   is just what is explicitly stated in A3, so that two assumptions are certainly
    consistent with each other. Two remarks follow. Firstly the state $| \nu_{\alpha} \rangle$
   is \underline{ not} produced, because there is no physical reason why it 
    should be. As already stated above only the mass eigenstates $| \nu_k \rangle$
    appear in physical amplitudes as initial or final states..
 As I have shown in a recent paper~\cite{JHF1}, assuming that such states
    are created in pion decay gives a prediction for the ratio $\Gamma(\pi \rightarrow  e \nu)/
  \Gamma(\pi \rightarrow \mu \nu)$ completely excluded by experiment. Interestingly enough,
    Giunti has himself recently given arguments based on quantum field theory why such
    states are {\it not} produced in pion decay~\cite{Giunti3}, wthout, apparently,
    realising the contradiction with the assumption A2 of TNO!
    Secondly, which 
    is essentially what is pointed out in the comment of Okun et al~\cite{Okun},
    the assumptions A2 and A3 of TNO are in clear contradiction with A4. On the one hand,
    it is assumed (A2 and A3) that the space-time velocities, $L/T$, are
    the same for all
     mass eigenstates, on the other hand (A4) they must have different kinematical 
     velocities, $p/E_k$. Aristotle made a wise remark that seems quite appropriate
     to this situation\footnote { On Giunti's website `Neutrino Unbound':
      http://www.nu.to.infn.it/ a fair number of wise quotations are to be found,
       even one in connection with a paper of mine. I therefore consider that I have 
       the right to include at least one here} 
       \par {\tt In inventing a model we may assume what we wish, but should avoid
        \newline impossiblities.}
     \par Since, however, the different mass eigenstates do not have to be produced at the same
      time in the different path amplitudes whose interference results in the 
       `neutrino oscillation' phenomenon, the hypotheses A2 and A3 are
        generally incorrect so that
      there in no contradiction with A4. I am, of course, using here Feynman's space-time
      formulation of QM which is the one best adapted to flavour oscillation problems,
      for the theoretical description of the problem.
      \par Thus Assumptions A2 and A3 are usually false. A coherent state  $| \nu_{\alpha} \rangle$
       is \underline{ not} produced~\cite{JHF1,Giunti3}, and although the source-detector distance,
      $L$, is the
      same for all states $| \nu_k \rangle$, the times-of-flight, $T$, (and hence the times
      of decay of the source) are different in the different amplitudes:
      $T_k/L = p_k/E_k$. Taking into account the different decay times of the 
       source increases the contribution of the neutrino propagators to the 
       interference phase by a factor of two and results in a numerically
       important contibution to the interference phase from the propagator
       of the source particle. There is therefore in fact no incompatiblity
       between the space-time and kinematical velocities, and no need, as asserted 
       by Giunti in TNO and Ref.~\cite{Giunti2}, to introduce an {\it ad hoc}
       gaussian `spatial wave-packet' to circumvent the problem. 
       \par There should be nothing shocking in the idea that a source can
        decay at different (but unobserved) times in quantum interference phenomena.
        Consider a Michelson interferometer with a 1m path difference between the two
        arms. The times of decay of the source atom that produces the photon
        that `interferes with itself' must obviously differ by 2m/c = 6.7 nsec
        when interference fringes are observed.
       \par Unfortunately, practitioners of the theory of neutrino oscillations
       appear to have little familiarity with Feynman's space-time approach to QM,
       being instead usually experts in text-book second-quantised field theory,
       which has
       limited relevance to space-time flavour oscillations. This is not the case,
       however, for atomic physics theorists. In concluding below I will briefly
       describe an atomic physics experiment that is an almost perfect analogue of
       the two-flavour neutrino oscillation problem. Feynman's path amplitude 
       method has been used to predict quantum interference effects in the 
       experiment in very good agreement with observation.
       \par One has the impression that neutrino physics theorists are stuck, for
        their conceptial understanding of QM, with Heisenberg in 1930. It is as 
        though Dirac's~\cite{Dirac} and Feynman's~\cite{Feyn}
       discoveries of the true role of space-time in QM
        were never made. At this very early period in the history of QM, statements
       like the following can often be found~\cite{Heis1}:
       \par {\tt There exists a body of exact mathematical laws but these 
         cannot \newline be interpreted as expressing simple relationships
       between objects \newline in space and time.}
       \par Or,~\cite{Heis2}:
         \par {\tt EITHER:}
       \begin{itemize}
       \item[ ]{\tt Causal relationship expressed by mathematical laws}
       \item[ ]{\tt BUT:$~~$physical description of space-time impossible}    
       \end{itemize}
       \par{\tt OR}
       \begin{itemize}
       \item[ ]{\tt Phenomena described in terms of space and time}
       \item[ ]{\tt BUT:$~~$ Uncertainty Principle}    
       \end{itemize} 
       The Feynman path amplitude formulation actually gives a mathematical law
       expressing a causal relationship (i.e. ordered in time) between
       elementary processes {\it in} space-time!
       The above quotations a very reminiscent of the woolly arguments invoked
       in blanket fashion by Giunti to avoid thinking clearly about the space
       time aspects of the QM of neutrino oscillations.
        \par For example, misuse of the Heisenberg Uncertanity relations~\cite{Giunti1}:
        \par {\tt The wave packet treatment of neutrino oscillations is also
       necessary for a correct description of the momentum and energy
       uncertainties necessary for the coherent production and detection of
       different massive neutrinos [12,13,9] whose interference generates the
       oscillations}
       \par It is not the wave functions of the neutrinos that are `coherent in the
        oscillation phenomenon, but rather the path amplitudes for the whole
        experiment. The neutrinos are only unobserved intermediate states in 
        these amplitudes.
        \par or~\cite{Giunti2}:
        \par{\tt One is free to define `so-called space velocities $\overline{v} 
          = x/t$ of massive \newline neutrinos which are trivially identical. However I
         think that such a \newline definition is useless. Neutrinos are not classical
         objects for which  \newline$v = x/t$. It is pretty clear that the uncertainty
         principle forbids any \newline  relation of this type}
         \par In fact, according to quantum field theory, on-shell
          particles {\it do} propagate classically  in space-time;
           only highly virtual ones show different
          behaviour. Notice the blanket invocation of `Heisenberg
          uncertainty' to block all further discussion. Actually the 
          relation $v = x/t$ is absolutely essential for the correcct 
          quantum mechanical description of neutrino oscillations.
         \par and~\cite{Giunti2}
          \par{\tt Of course energy and momentum are not exactly defined
           in the production process, in order to allow the coherent production
            of different massive \newline neutrinos, but it is pretty unlikely that
          the uncertainty in energy and \newline momentum could generate the equal energy
          constraint.}
          \par In relativistic quantum field theory energy and momentum
           are both exactly conserved at all vertices. `Heisenberg uncertainty'
           (see below) only appears as a smearing of the masses of
           virtual particles about pole values. Another, desperate, last resort
           invocation of the `Uncertainty Principle'!
         \par As discussed in detail in my paper Ref.~\cite{JHF2}, the misuse 
         of spatial wavepackets by Giunti and many other authors is related to
       a confusion of QM with classical wave theory, for which the concepts of
       phase velocity, group velocity and wave packets are useful and meaningful
      concepts. In the Dirac-Feynman formulation of QM they become almost
      irrelevant. Still, the Heisenberg Uncertainty Principle is an important
      part, when correctly used, of QM. I describe below its application to
      neutrino production in pion decay. In fact the neutrinos in
       $\pi \rightarrow \mu \nu$ (not in  $\pi \rightarrow e\ \nu$) 
       are described by a momentum wave packet. It has however nothing to do 
       with the Fourier transform of the arbitary gaussian spatial function that is
       parachuted into the theory by Giunti and many other authors.

       \par It is a consequence of quantum field theory
        that on-shell particles, or virtual particles propagating over macroscopic
        spatial separations,
        do so in a classical manner~\cite{Mohanty}. The dominant trajectories 
        in Feynman path integrals are the classical ones. Consider the 
        space-time description of pion decay, $\pi \rightarrow \mu \nu$,
          in QM. The Heisenberg Uncertainty
        Principle (so dear to the hearts of neutrino physics theorists!)
        should be correctly invoked in three different ways in the description
        of this process: (i) the decay width $\Gamma$ and the decay lifetime
        $\tau$ certainly respect the energy-time Uncertainty Relation
        $\Gamma \tau =1$; (ii) as a result of the finite pion lifetime
        the physical mass of the pion differs from the pole mass. 
        In accordance with the energy-time Uncertainty Relation
        the mass smearing, given by a Breit-Wigner amplitude, is
        inversely proportional to the decay lifetime; (iii) due to the
        finite muon decay lifetime, the physical mass of the muon   
        is also smeared around its pole value. Since the muon is
        unobserved, overall energy-momentum conservation leads to a neutrino
        momentum wave-packet in the case of $\pi \rightarrow \mu \nu$
        but not for $\pi \rightarrow e\ \nu$, since the electron has 
        an infinite lifetime. The effects of this wave packet are
         calculated in my paper~Ref.~\cite{JHF3}. They are
        tiny. The upshot is, that once the neutrinos are produced
        (respecting, of course, the law  $\Gamma \tau =1$) they propagate
        in space-time as essentially classical particles. This wave 
        packet is the only one that is relevant to pion 
        decay. In particular there is no {\it ad hoc} gaussian spatial
       wave packet describing the source. The uncertainty in the position
       of the source particle, a property of the initial state, and therefore
       the same in all path amplitudes, gives only a small, incoherent,
        correction to the oscillation phase. Again, this effect is calculated
       in Ref.\cite{JHF1} and found to be very small.
        \par  I would like to remark that I do agree with Giunti that
       the `equal energy' hypothesis proposed in Ref.~\cite{Okun} is not
       relevant to resolving the evident contradiction between assumptions
       A2 and A3 of TNO as compared to A4. As shown in Ref.~\cite{JHF2} sufficient
       assumptions to obtain the standard oscillation phase are A2 or A3
       (since A2 implies A3). The result at, O($m_{\nu}^2$), is independent of the kinematical
        assumptions such as A4 (equal momenta, equal energies or exact energy-momentum
        conservation). In fact, in the papers by Stodolsky~\cite{Stodolsky} and 
         Lipkin~\cite{Lipkin} cited ini Ref.~\cite{Okun}, not onlt is 
      the equal energy assumption made, but the temporal part of the 
      neutrino propagator phase (necessarily non-zero in the laboratory 
       system by Lorentz invariance) is, arbitarily and incorrectly, set to zero.
        \par  A misnomer which occurs widely in Refs.~\cite{Giunti1,Okun,Giunti2}
         is `plane wave' for `space-time propagator'. This function has meaning
        only as a factor in a path amplitude. Interpreted as a wavefunction 
        it is non square-integrable and therefore devoid of any physical 
        significance.
            
         \par The atomic physics experiment that provides a close analogy with
          a two flavour neutrino oscillation problem is called the 
          `Photodetachment Microscope~\cite{PM}.A coherent source of electrons
         of fixed energy is provided by a negative ion beam irradiated by a laser.
         The detached electron moves in a constant electric field before detection.
          Just two classical trajectories link the point of emission to any
          point on a plane detector oriented perpendicularly to the 
          electric field direction. Quantum interference effects are observed
          between the path amplitudes corresponding to the two trajectories.
          A good pedagogical description can be found in Ref.~\cite{BBD} 
          where the appropriate path integral formula\footnote{ A similar formula has
        recently been proposed for the neutrino oscillation problem by Pa\u{z}ma and Vanko
         ~\cite{PV}. The corresponding oscillation phase was not, however, derived.}r 
: 
  \begin{equation}
  \psi(\vec{r},t_f)  = \int_{-\infty}^{t_f}\exp[i\frac{\epsilon t_i}{\hbar}]
      \exp[i\frac{S_{cl}(\vec{r},t_i,t_f)}{\hbar}]d t_i \nonumber
   \end{equation}
    is given.
  \par In this formula $\epsilon$ is the energy of the detached electron and
    $S_{cl}$ the classical action corresponding to an electron
    trajectory. Note particularly the time integral on the
    RHS of the equation. The first exponential function is the propagator of
    the coherent source (analagous to that of a coherent neutrino source) the
    second represents the propagator of the electron in the electric field. 
    In practice it is well approximated by the contributions of the two classical
    trajectories mentioned above, corresponding to values of
    $t_i$ with a fixed separation. These are the analogues of the propagators
    of different neutrino mass eigenstates. A typical value of the difference
    in $t_i$ between the two trajectories, quoted in Ref.~\cite{BBD} is
  160 psec for a time-of-flight of 117 nsec. The laws of physics are the same
  in this and in any neutrino oscillation experiment. Contrary to the claims in
   TNO and Ref.~\cite{Giunti2}], no {\it ad hoc} `wave packets' are required
   for a correct quantum mechanical description of such systems. The standard
   formula for the oscillation phase which, as discussed in Ref.~\cite{JHF2}
   may well be correct for heavy quark oscillations, is then in most cases,
   not so~\cite{JHF1}
   for neutrino oscillations.
   \par Finally, it is remarked that as described in Ref.~\cite{JHF1} a clear
   experimental discrimination between the standard and path amplitude formulae
   of the oscillation phase is provided by long-linebase `$\nu_{\mu}$
   disappearence' experiments with pion and kaon source beams.

\pagebreak

\end{document}